\documentclass[aps,twocolumn,prl,amsmath,caption,amssymb,superscriptaddress,showpacs]{revtex4-1}
\usepackage{graphicx}
\usepackage[separate-uncertainty=true]{siunitx}
\newcommand{\ppf}{\emph{Physarum polycephalum}}
\newcommand{\pp}{\emph{P.~polycephalum}}

\begin{document}

\title{Fluid flows shaping organism morphology}

\author{
Karen Alim}

\address{Max Planck Institute for Dynamics and Self-Organization, Am Fassberg 17, D-37077 G\"ottingen, Germany}



\email{karen.alim@ds.mpg.de}

\begin{abstract}
A dynamic self-organized morphology is the hallmark of network-shaped organisms like slime moulds and fungi. Organisms continuously re-organize their flexible, undifferentiated body plans to forage for food. Among these organisms the slime mould \pp\ has emerged as a model to investigate how organism can self-organize their extensive networks and act as a coordinated whole. Cytoplasmic fluid flows flowing through the tubular networks have been identified as key driver of morphological dynamics. Inquiring how fluid flows can shape living matter from small to large scales opens up many new avenues for research. 
\end{abstract}
\maketitle
Many organisms, including a broad range of species of slime moulds and fungi, grow and forage as a single large network. Networks change their morphology over and over again during growth and migration to locate food in a patchy environment \cite{Boddy:09,Stephenson:94}, see also Fig.~1. Moreover, network morphology adapts in response to newly acquired food sources, even connecting food sources in an efficient and robust manner \cite{Tero:10}. The striking similarity of the morphological dynamics in foraging fungi and slime moulds is even more surprising if one takes into account that slime moulds and fungi are genetically distinct, with slime moulds being genetically even closer to animals than fungi. It is therefore likely that not biological make-up but the physics of fluid flows within the tubular networks are critical to the self-organization of network morphology across an individual. Both kinds of living, adaptive networks exhibit oscillatory, long-ranged fluid flows \cite{Tlalka:02,Tlalka:07,Kamiya:50}. Here, the syncytial plasmodia of the slime mould \ppf\ emerged as a model system to understand the role of flows in coordinating morphology. Fluid flows in this organism are highly coordinated, driving intracellular transport on short time scales but also migration and likely morphological self-organization at long time scales.

\begin{figure*}[t!]
\centering
\label{fig_morpho}
\includegraphics[width = \textwidth]{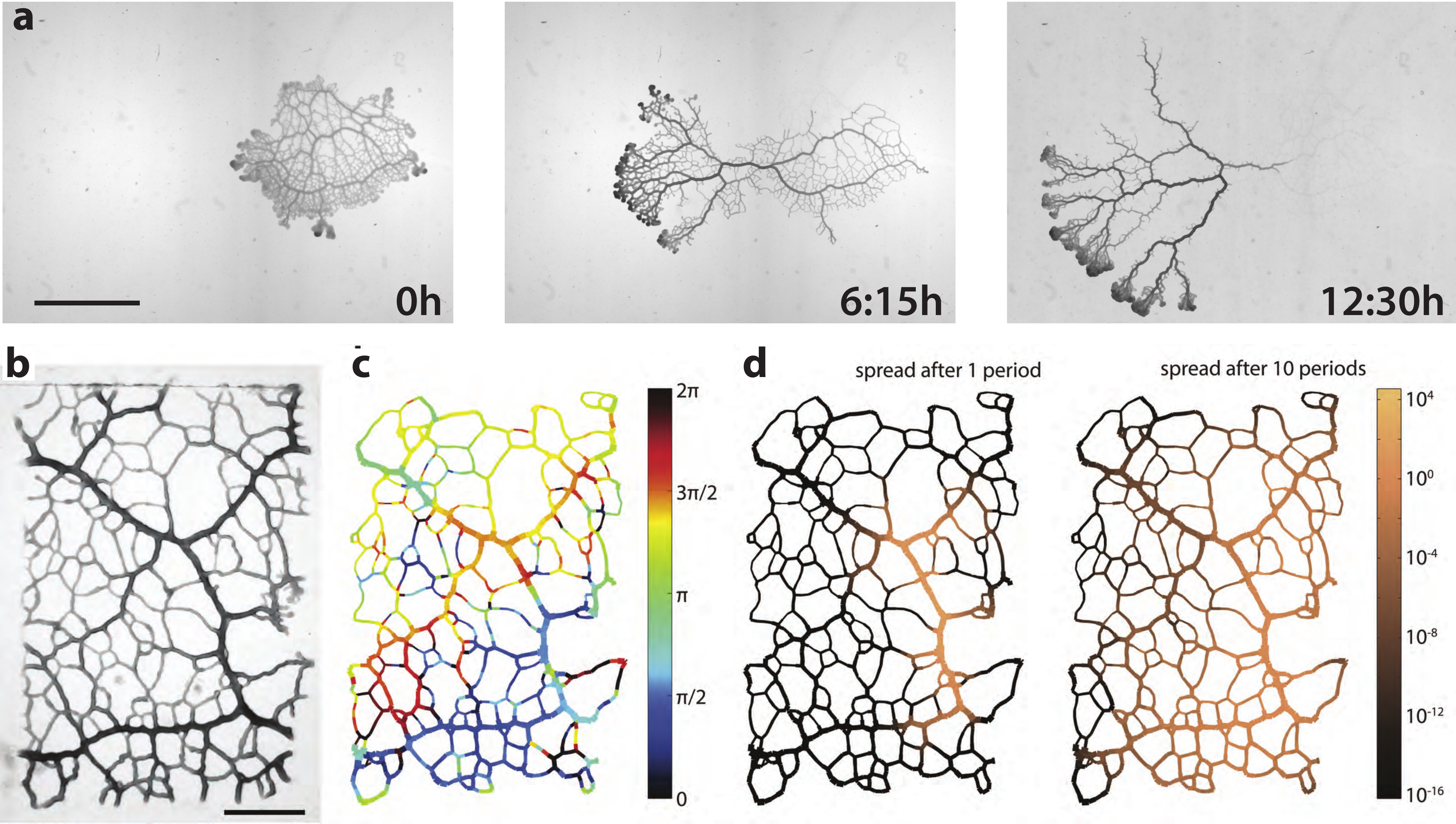}
\caption{
Network morphology and peristaltic wave driving long-ranged cytoplasmic flows in the plasmodial slime mould \pp\. (a) Self-organization of the network morphology over the course of 12.5 h. An initially well-reticulated network migrates to the left thereby retracting connections and altering tube radius hierarchy. Scale bar, \SI{5}{mm}. (b) Bright-field microscopy image of P. polycephalum network trimmed to a rectangular shape. Scale bar, \SI{2}{mm}. (c) Analysis of actin-cytoskeleton-driven contractions of tube walls in (b) reveal phases of contractions to be patterned in a roughly linear gradient. The wavelength of this peristaltic phase pattern on a network matches the organism size. (d) Simulation results of the spread of particles (copper colour) due to the streaming of fluid flow driven by the peristaltic contractions.
}
\end{figure*}
\section{Cytoplasmic flows organized in peristaltic wave}
In \pp\ the fluid cytoplasm within the tubular network streams forth and back in a shuttle flow \cite{Stewart:59,Kamiya:50}. Network sizes are macroscopic ranging from about \SI{500}{\mu m} to \SI{0.5}{m} - experiments are typically conducted on specimen of up to few \SI{}{cm} in size. Flows generally exhibit a Poiseuille profile \cite{Kamiya:50,Bykov:2009} with deviations likely in smaller tubes \cite{Romanovskii:1995}. Flow is dominated by small Reynolds number $\mathrm{Re} = 2U R/\nu \approx 0.002$ and small Womersley number $\alpha=R\sqrt{\omega/\nu}\approx 0.004$, based on a representative tube radius of $R=\SI{50}{\mu m}$, a flow velocity reaching up to $U=\SI{1}{mm/s}$ \cite{Kamiya:50}, a kinematic viscosity of cytoplasm $\nu = \SI{6.4E-6} {m^2/s}$ \cite{Swaminathan:97} and an oscillation frequency of $\omega=\SI{0.05} {Hz}$. The cytoplasm is enclosed by gel-like walls that are lined with an actin cortex \cite{Isenberg:76,NaibMajani:1983uo}. Actin organizes in circumferential fibrils that contract periodically  \cite{Rieu:2015jr} and drive the cytoplasms' shuttle flow. Contrary to long-lasting speculations about localized pumps driving pressure difference, the common understanding now is that flows arise through network-wide, self-organized contractions of the actin cortex \cite{Alim:13,Teplov:2017}. 

The shuttelling cytoplasm itself is very rich in actin. In \SI{1}{mm} sized cytoplasm extract droplets, so called proto-plasmic droplets, the actin cortex and the contractions self-organize over time, showing first irregular contraction patterns and later highly coordinated spatio-temporal patterns including standing, travelling and spiral waves \cite{Matsumoto:88,Ueda:05,Takagi:08,Takagi_PhysicaD_2010}. Similar dynamics and patterns can be reproduced in models by describing the proto-plasmic droplets by two phases of a viscoelastic solid phase representing the cytoskeleton, interpenetrated by a second fluid phase representing the cytosol and coupling both phases by a soluble molecule that activates tension in the solid phase \cite{Radszuweit:2013, Alonso:2016}.  Even on the much larger scale of an entire tubular network contractions are also highly coordinated, forming a peristaltic wave spanning specimen of at least \SI{2}{cm} in size \cite{Nakagaki:2000vk,Alim:13},  see Fig.~1. While the contraction period only increases moderately with network size \cite{Kuroda:2015ff,Ueda:1982tr}, the wavelength of the peristaltic wave matches organism size spanning two orders of magnitude \cite{Alim:13}. It is fascinating to investigate how this scaling can arise.  Given the success of mechanochemical models for proto-plasmic droplets it is likely that also the organism spanning peristaltic wave is a result of mechanochemical patterning, which could nicely complement our emerging understanding of this novel patterning mechanism widespread in biological systems  \cite{Howard:2011,Gross:2017}. From a fluid mechanics point a view a peristaltic wave matching organism size induces the highest flow velocities throughout a network. As flows change with organism morphology, they are likely not only important for organism homoeostasis but also for the coordination of morphological adaptation itself.

\section{Effective intracellular transport by oscillatory flow}
Peristalsis is a common mechanism in biological systems, creating oscillatory flow and pumping fluid along a tube \cite{Shapiro:69,SHAPIRO:71,Selverov:2001ty,Li:2006jf}. In \pp\ the tubular network can be considered to be of fixed volume on the time scales of tens of contraction periods. Therefore, net fluid transport is not relevant on these short time scales \cite{Iima:2012jg,Alim:13}. Yet, creating shuttle flows by a peristaltic wave of contractions is a simple but powerful mechanism to increase the spread of any particles, like metabolites or signalling molecules within this closed network. Rare measurements show organism-wide $\sim \SI{2}{cm}$ transport of particles within half a contraction period \cite{Nakagaki:2000vk} out-competing diffusive spread that would have travelled only $\SI{0.25}{mm}$ in that time frame. Because of the oscillatory nature of flow, particles flow mainly back to their initial site after a whole period. But the peristaltic flow also increases the effective diffusion $\kappa\rightarrow\kappa+U^2R^2/48\kappa$ according to Taylor dispersion in long slender tubes \cite{Aris:56,Taylor:53} also applicable to contractile tubes \cite{Mercer:90,Mercer:94}. Here, $\kappa$ denotes the bare molecular diffusivity. Rapid diffusion across the tiny tube cross-sections allows particles to transition between fast and slow streamlines of the Poiseuille profile rapidly increasing their dispersion along the tube by an order of magnitude, see Fig.~2 c. The adjacency of big and small tubes and therefore different flow velocities in a slime mould network make it a hard task to theoretically map out how far particles can spread \cite{Heaton:2012}. One successful strategy is to map out an effective dispersion \cite{Saffman:59,Marbach:2016bf}, which unveils for example that particle spread is increased by pruning/coarsening of the network when \pp\ is left to starve \cite{Marbach:2016bf}. This already provides a glimpse of the challenging question on how network morphology impacts network-wide transport \cite{Nakagaki:2000vk}. Is morphology geared to optimizing dispersion? If so, how does it evolve toward an optimized morphology? Transport by fluid flow seems to lie at the basis of network self-organization since advected signaling molecules may propagate information about the acquisition of a food source throughout the network by hijacking fluid flows \cite{Alim:2017fs}. Here, signaling molecules advected by fluid flow directly increase contraction activity. Further, flow is necessary to synchronize and coordinate contractions \cite{Yoshimoto:1978wg,Samans:84,Achenbach:81}. It is therefore likely that flows play a crucial role in coordinating contractions over space and time. 
\section{Cell migration by pumping of peristaltic wave}
Cytoplasmic flows form the basis of fast locomotion of very different kinds of amoeba \cite{Taylor:1973jr,Lammermann:2008ib,Charras:2008gf,Yoshida:2006bk}. Flows arise by local expansion of the actin cortex and subsequent myosin dependent contraction \cite{Lammermann:2008ib}. Flows and contractility underlying cell migration are well accessible in the amoeboid plasmodia of \pp. Plasmodia of \SIrange{100}{500}{\mu m} length adopt an amoeboid shape migrating rapidly \cite{Matsumoto:08,Zhang:2017vx} by net fluid transport generated by a contractile peristaltic wave and cortex expansion at the front \cite{Lewis:2015}. 
Changes in contraction pattern affect locomotion velocity \cite{Rodiek:2015hj}. Plasmodia that exceed $\sim \SI{500} {\mu m}$ form a full network structure often including multiple migration fronts. Over the peristaltic cycle a front location advances and retracts asymmetrically leading to net advancement at long time scales \cite{Baumgarten:2014bv}. As networks grow they move faster \cite{Kuroda:2015ff}. When confined to lanes, migration velocity scales linearly with maximal plasmodium height, reaching locomotion speeds of up to $v=0.4mm/s$ \cite{Kuroda:2015ff}. Reorientation of migration direction for example toward a food source is associated with a redirection of the peristaltic wave direction to that site \cite{Alim:2017fs} further substantiating that migration is governed by fluid flows on long time scales. Given that signaling molecules are advected by fluid flows affecting actin cortex dynamics \cite{Alim:2017fs} it is seems possible that flows play an important role in the navigation of organisms, acting for example during chemotaxis.
\section{Morphological changes triggered by cytoplasmic flows}
Fluid flows not only transport particles and fluid mass but also exert forces themselves that may induce long-term changes to morphology. Forces may directly feed back onto biochemical reactions triggering complex spatiotemporal dynamics due to this mechanochemical coupling as currently more and more observed in morphogenetic processes \cite{Howard:2011,Gross:2017}. Even without the ability to pin down a specific feedback on chemical reactions the influence of forces generated by flow can be investigated on a coarse-grained level. For example, in animal vasculature it is observed that tube diameters grow with increased flow rate regulating shear force to a balanced level \cite{KAMIYA:1980wg}. This observation inspired the idea that fluid shear force induces morphological changes in vasculature, a concept also successfully used in models of \pp\ dynamics \cite{Nakagaki:2000vk}. Further support is the success of Murray's law particularly in plant and animal vasculature \cite{Murray:1926tj} which predicts the ratio of tube diameters at a network node under the assumption of conserved shear force. Murray's law is also consistent with minimizing dissipation inspiring theoretical work on optimal network architectures \cite{Corson:10,Katifori:10,Durand_PRL_2007}. Given that \pp\ grows its almost transparent tubes in a planar network, testing principles such as Murray's law \cite{Akita:2016fp} and in general relating morphological dynamics to flow properties is very feasible. In particular, the adaptability of the network morphology makes it a very suitable system to explore how well certain properties like dissipation, robustness \cite{Tero:10} or transport capabilities \cite{Meigel:2017,Chang:2015} are optimized by living organisms. Equally, as flows are globally coupled throughout the network, there is considerable additional complexity in this system. Indeed, it might well be that precisely this added complexity due to the coupling is the key to have simple mechanisms based on fluid flow give rise to the complex dynamics of self-organization of morphology we observe.
\section*{Acknowledgement}
This work has been supported by the Max Planck Society

\end{document}